\documentclass{aa} 
\usepackage{natbib}
\usepackage{epsfig,latexsym,graphicx,epsf,subfigure}
\usepackage{threeparttable,hhline,longtable}
\usepackage{txfonts}

\newcommand{\eqref}[1]{(\ref{#1})}

\def\lsim{\raise0.3ex\hbox{$<$}\kern-0.75em{\lower0.65ex\hbox{$\sim$}}}
\def\gsim{\raise0.3ex\hbox{$>$}\kern-0.75em{\lower0.65ex\hbox{$\sim$}}}

\begin{document}
\title{The intrinsic colour dispersion in Type~Ia supernovae}

\author{S.~Nobili\inst{1} \and
        A.~Goobar\inst{1} \and 
        R.~Knop\inst{2}  \and 
        P.~Nugent\inst{3} }
\authorrunning{Nobili et al.}

\offprints{S.~Nobili , serena@physto.se}

\institute{Department of Physics, Stockholm University, \\
         SCFAB, S--106 91 Stockholm, Sweden \and
Department of Physics \& Astronomy, \\Vanderbilt University, USA \and  
Lawrence Berkeley National Laboratory, \\
         1 Cyclotron Road, Berkeley, CA 94720, USA}

\date{Received ...; accepted ...}

\date{} \abstract{ The properties of low-redshift Type Ia supernovae
are investigated using published multi-band optical broadband data
from the Calan/Tololo and CfA surveys. The average time evolution of
$B-V$, $V-R$, $R-I$, $B-I$ and $V-I$, the intrinsic dispersion and
time correlations are studied. This information is required to deduce
the extinction of such explosions from the measured colours. We find
that extinction corrections on individual SNe based on their colours
up to 40 days past the ${\mathit B}$-band lightcurve maximum are generaly limited
to $\sigma_{A_V} \gsim 0.1$, due to intrinsic variations, as far as it
can be conservatively deduced with the current sample of data.
However, we find that the $V-R$ colour, especially at late times, is
consistent with a negligible intrinsic spread, and may be the most
accurate estimator for extinction. } 
\maketitle

\section{Introduction}

In the last few years Type~Ia supernovae (SNe~Ia) have proved to be
excellent distance estimators and have been successfully used to
investigate the fate of the universe \citep{Perl99,Riess}. Despite the
broad use of these objects by cosmologists, the current knowledge of the nature
of SNe~Ia is rather limited. Thus, there is a strong demand for further
understanding to assess important issues about these explosions. For
cosmological implications, the main concerns are related to the possible evolution
of the SN properties with redshift. Moreover, critical tests for
extinction along the line of sight based on supernova colours require
good knowledge of the intrinsic properties of these objects.

In this paper, a statistical study on 48 well observed
nearby SNe~Ia is carried out. In particular, the intrinsic dispersion
in SN colours is investigated using published ${\mathit BVRI}$ data. We also focus on the time
correlation of intrinsic optical colours.  This information is needed
to address the possible host galaxy or intergalactic extinction by dust of
supernovae used for cosmological tests (see
e.g. \cite{riess99q,beethoven}), and also to probe for other exotic
sources of dimming at high-z with differential extinction \citep{axion}.

\section{The data set}

\begin{figure}[htb]
\includegraphics[width=4.1cm]{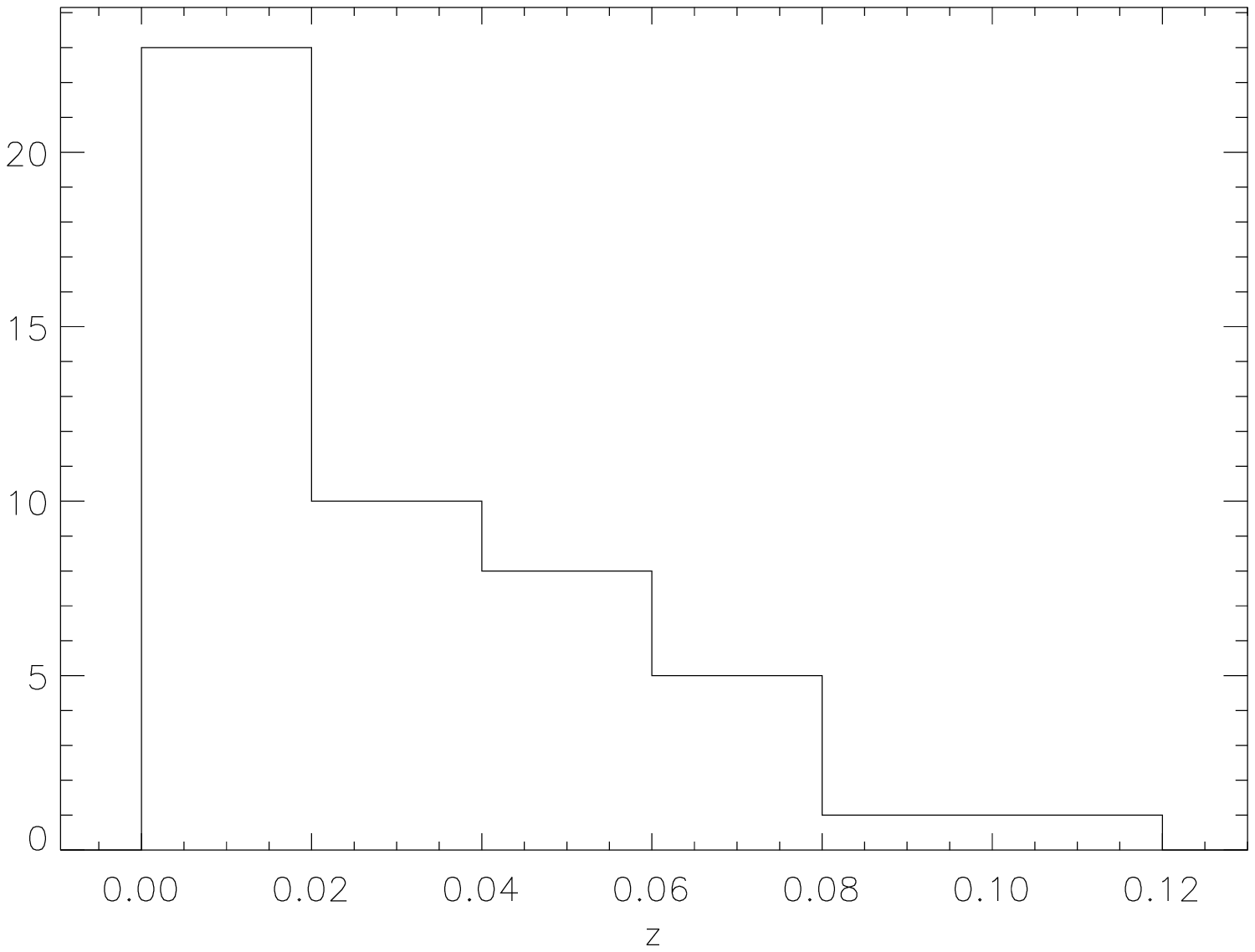}
\includegraphics[width=4.1cm]{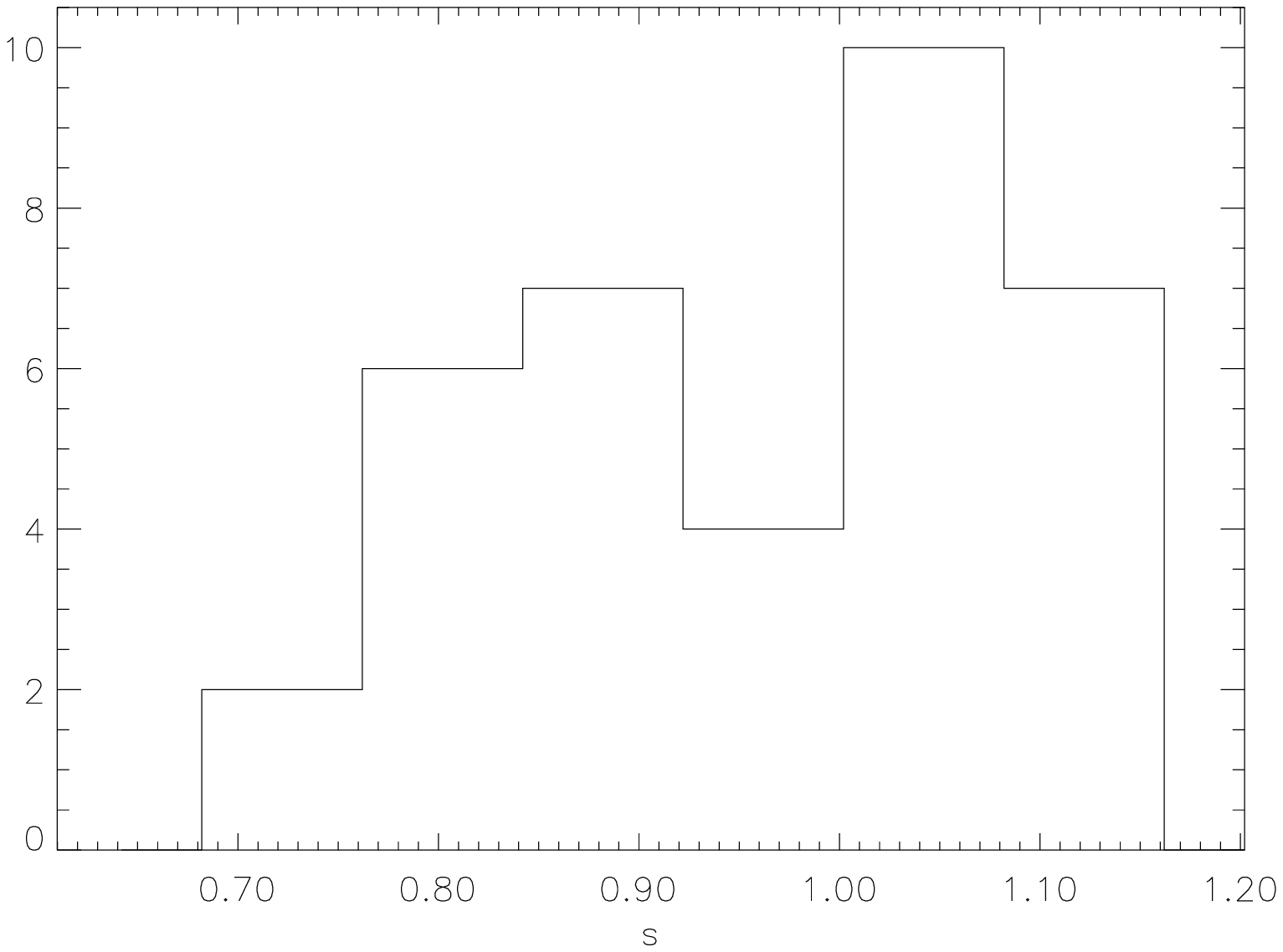}
\caption{ Distribution of redshift, {\it z}, (binsize=0.02), and
stretch factor, {\it s}, (binsize= 0.08), for the analysed sample}
\label{zs_histo}
\end{figure}

Published ${\mathit BVRI}$ lightcurves of well observed nearby SNe~Ia were
analysed. The considered sample consists of 48 SNe~Ia from 2 different
sets, the Calan/Tololo data published by \citet{Hamuy}, and the set in
\citet{Riess22}, usually referred to as the CfA data. The list of SNe
is given in Table~\ref{listSNe}, along with the observed filter data
available for each of them, their redshift and the ${\mathit B}$-band lightcurve
``stretch'', {\it s}, as defined in \citet{Perl97} and
\citet{goldhaber}.\\ The selected samples include a broad variety of
SNe, well distributed in stretch factor, {\it s}. This parameter,
related to $\Delta m_{\rm 15}$ \citep{deltam15}, has been found to
correlate with the supernova luminosity. Thus, a sample well
distributed in stretch should imply a broad distribution in
luminosity. Fig.~\ref{zs_histo} shows the distribution of the ${\mathit B}$-band
light curve stretch factor and SN redshifts ($0.003 \le z \le 0.12$),
for both samples. The timescale stretch parameter was determined from
the lightcurve fits as in \citet{goldhaber} using their ${\mathit B}$-band
lightcurve template with a parabolic behavior for the earliest epoch
after explosion.

$K$-corrections were applied to account for the small cosmological
redshift as described in e.g. \citet{snoc}, using the ($s=1$)
spectroscopic template of \citet{Nugent-kcorr} as a starting point.
The results of this analysis were used to improve the spectral
template of SN~Ia's (Sect.~\ref{sec:template}) and we iterated the
analysis once re-calculating the $K$-corrections with the improved
template. Note that even though the $K$-corrections for the 
used data set are small, typically 
of the order of a few hundreds of a magnitude, for some of the more
distant objects in the sample considered, they reach up 
to $\sim 0.5$ mag.

The SNe light curves were corrected for both Milky Way and
host galaxy extinction as in \citet{Phillips99} using the method first
proposed by \citet{Lira} for estimating host galaxy extinction using
late epoch light curves. There is empirical evidence that the $B-V$
colors of SNe~Ia show extremely small scatter for the period between
30 and 90 days post ${\mathit B}$-band maximum, despite any difference in the
light curve shapes at earlier epochs. As the set of SNe used in this
article is a subset of the one analysed in \citet{Phillips99}, the
host galaxy extinctions listed in Table 2 of their paper were used.
These were derived combining the late time $B-V$ colour with information
on the $B-V$ and $V-I$ at maximum light.

The \citet{Cardelli89} relation, modified by \citet{Odonnell}, was
used to compute the extinction in other colours given $E(B-V)$.
Spectral templates of Type~Ia SNe were used to compute the evolution
of the extinction with the supernova epoch.

\begin{figure}[htb]
\includegraphics[width=7cm]{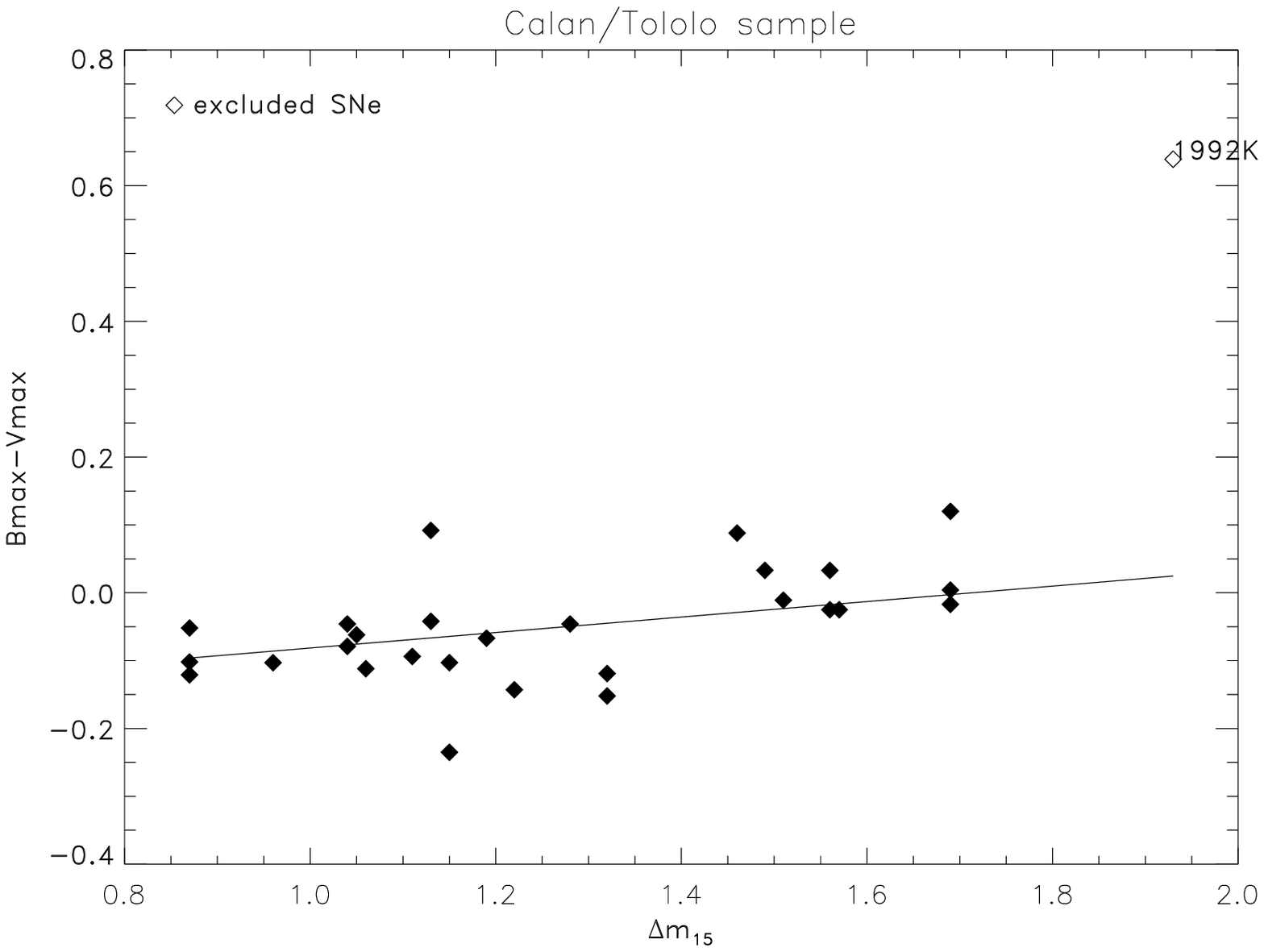}
\includegraphics[width=7cm]{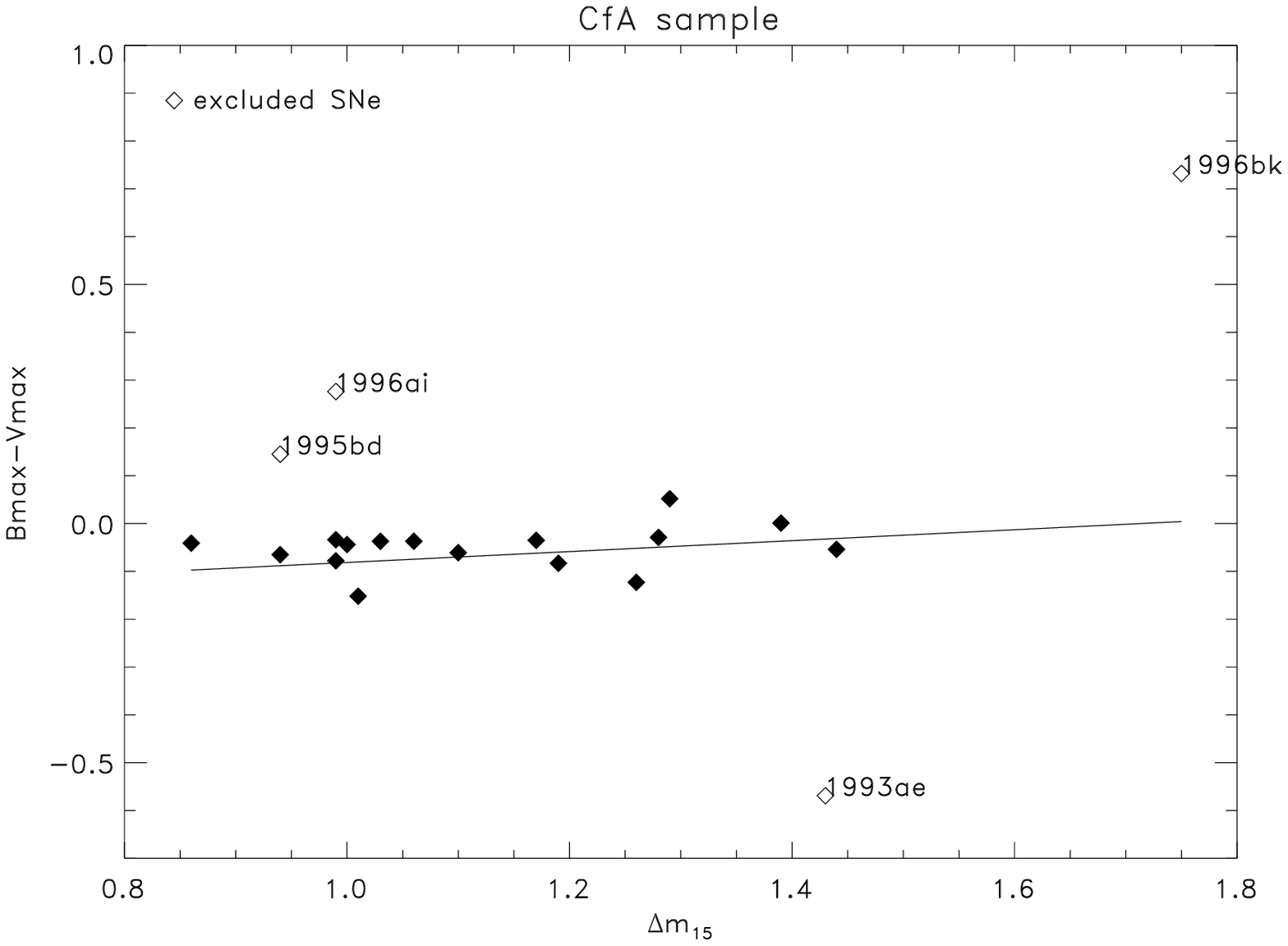}
\caption{$B_{\rm max}$-$V_{\rm max}$ vs $\Delta m_{\rm 15}$ for the 2
sets of data, k-corrected, corrected for Galactic extinction and Host
Galaxy extinction, compared with the result of \citet{Phillips99}}
\label{BmaxVmax}
\end{figure}

The extinction corrected lightcurves were further screened to exclude
the most peculiar SNe, as the main emphasis of this work is to
establish the properties of ``normal'' supernovae. Fig.~\ref{BmaxVmax}
shows the difference of the $B$-band lightcurve maximum, $B_{\rm max}$
and the $V$-band light curve maximum $V_{\rm max}$ plotted against the
decline rate parameter, $\Delta m_{\rm 15}$, as reported in
\citet{Phillips99}. A $3\sigma$ clipping rejection criteria was
applied, iterating until no data points were further rejected. At
least 5 SNe in the 2 data sets deviate significantly from the expected
linear relation derived in \citet{Phillips99} from an independent set
of 20 non-reddened SNe (solid line in Fig.~\ref{BmaxVmax}):
$$B_{\rm max}-V_{\rm max}=-0.07(\pm 0.012)+0.114(\pm 0.037)(\Delta
m_{\rm 15}-1.1)$$

The outliers are SN~1993ae, SN~1995bd, SN~1996ai, and SN~1996bk for
the CfA data, and SN~1992K for the Calan/Tololo set. Moreover SN~1995E
and SN~1995ac have been excluded since they are similar to the
peculiar SN~1991T \citep{bfn93}.\footnote{Including these 2 SNe in
the analysis does not change the results significantely.}

\begin{table}[htb]
\caption[table SNe]{List of SNe used for the analysis.\\(1),
\citet{Riess}; (2), \citet{Hamuy}; \\$^{a}$ excluded from the
analysis presented here.}
\begin{center}
\begin{tabular}{lllll}
\hline
\hline
 SN & band &  $z$ & $s$ & Ref. \\
\hline
1993ac &  B,V,R,I & 0.049 & 0.865 & (1)\\
1993ae$^{a}$ &  B,V,R,I & 0.019 & 0.846 & (1)\\
1994ae &  B,V,R,I & 0.004 & 1.033 & (1)\\
1994M  &  B,V,R,I & 0.023 & 0.865 & (1)\\
1994Q  &  B,V,R,I & 0.029 & 1.116 & (1)\\
1994S  &  B,V,R,I & 0.015 & 1.061 & (1)\\
1994T  &  B,V,R,I & 0.035 & 0.890 & (1)\\
1995ac$^{a}$ &  B,V,R,I & 0.050 & 1.123 & (1)\\
1995ak &  B,V,R,I & 0.023 & 0.857 & (1)\\
1995al &  B,V,R,I & 0.005 & 1.044 & (1)\\
1995bd$^{a}$ &  B,V,R,I & 0.016 & 1.131 & (1)\\
1995D  &  B,V,R,I & 0.007 & 1.081 & (1)\\
1995E$^{a}$  &  B,V,R,I & 0.012 & 1.024 & (1)\\
1996ai$^{a}$ &  B,V,R,I & 0.003 & 1.110 & (1)\\
1996bk$^{a}$ &  B,V,R,I & 0.007 & 0.761 & (1)\\
1996bl &  B,V,R,I & 0.036 & 1.030 & (1)\\
1996bo &  B,V,R,I & 0.017 & 0.902 & (1)\\
1996bv &  B,V,R,I & 0.007 & 1.106 & (1)\\
1996C  &  B,V,R,I & 0.030 & 1.102 & (1)\\
1996X  &  B,V,R,I & 0.007 & 0.889 & (1)\\
\hline
\hline
1990af &  B,V     & 0.051 & 0.792 & (2)\\
1990O  &  B,V,R,I & 0.030 & 1.116 & (2)\\
1990T  &  B,V,R,I & 0.040 & 0.998 & (2)\\
1990Y  &  B,V,R,I & 0.039 & 1.007 & (2)\\
1991ag &  B,V,R,I & 0.014 & 1.084 & (2)\\
1991S  &  B,V,R,I & 0.055 & 1.114 & (2)\\
1991U  &  B,V,R,I & 0.032 & 1.068 & (2)\\
1992ae &  B,V     & 0.075 & 0.970 & (2)\\
1992ag &  B,V,I   & 0.025 & 0.951 & (2)\\
1992al &  B,V,R,I & 0.015 & 0.963 & (2)\\
1992aq &  B,V,I   & 0.102 & 0.868 & (2)\\
1992au &  B,V,I   & 0.061 & 0.787 & (2)\\
1992bc &  B,V,R,I & 0.020 & 1.039 & (2)\\
1992bg &  B,V,I   & 0.035 & 0.983 & (2)\\
1992bh &  B,V,I   & 0.045 & 1.048 & (2)\\
1992bk &  B,V,I   & 0.058 & 0.825 & (2)\\
1992bl &  B,V,I   & 0.044 & 0.812 & (2)\\
1992bo &  B,V,R,I & 0.019 & 0.767 & (2)\\
1992bp &  B,V,I   & 0.079 & 0.907 & (2)\\
1992br &  B,V     & 0.088 & 0.682 & (2)\\
1992bs &  B,V     & 0.064 & 1.030 & (2)\\
1992J  &  B,V,I   & 0.045 & 0.798 & (2)\\
1992K$^{a}$  &  B,V,I   & 0.010 & 0.787 & (2)\\
1992P  &  B,V,I   & 0.037 & 0.952 & (2)\\
1993ag &  B,V,I   & 0.049 & 0.917 & (2)\\
1993B  &  B,V,I   & 0.070 & 1.023 & (2)\\
1993H  &  B,V,R,I & 0.024 & 0.774 & (2)\\
1993O  &  B,V,I   & 0.051 & 0.950 & (2)\\
\hline
\end{tabular}
\label{listSNe}
\end{center}
\end{table}

\section{The average optical colours}
\label{sec:average  }

\begin{figure}[htb]
\includegraphics[width=8cm]{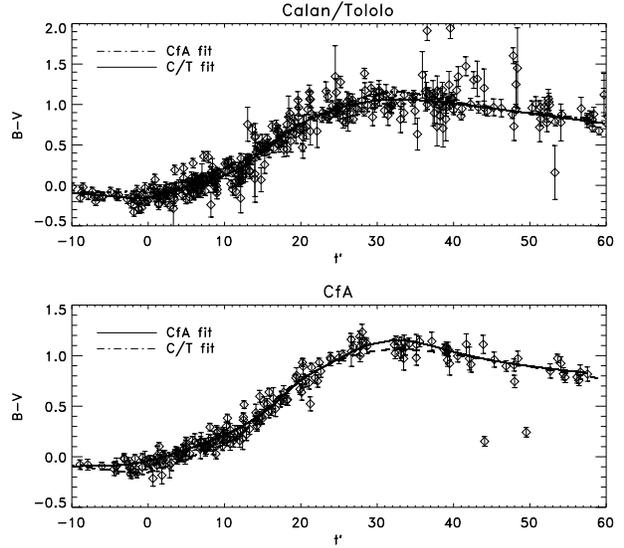}
\caption{$B-V$  for the two sets of data. The
solid line represent alway the curve found for the current set, while
the dashed line is the curve found for the same colour of the other
set.}
\label{coloursBV}
\end{figure}

\begin{figure}[htb]
\includegraphics[width=8cm]{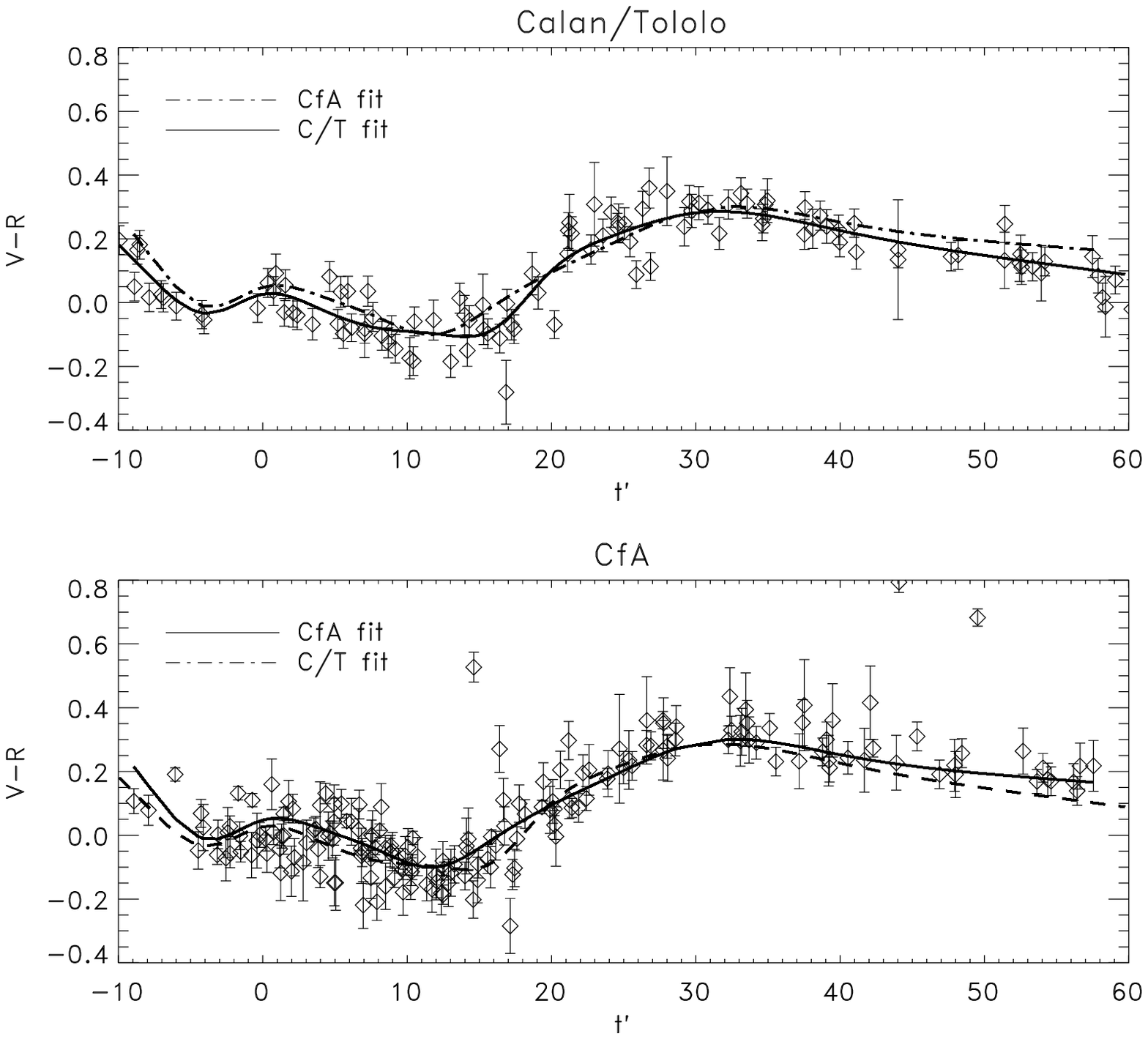}
\caption{$V-R$  for the two sets of data. The
solid line represent alway the curve found for the current set, while
the dashed line is the curve found for the same colour of the other
set.}
\label{coloursVR}
\end{figure}

\begin{figure}[htb]
\includegraphics[width=8cm]{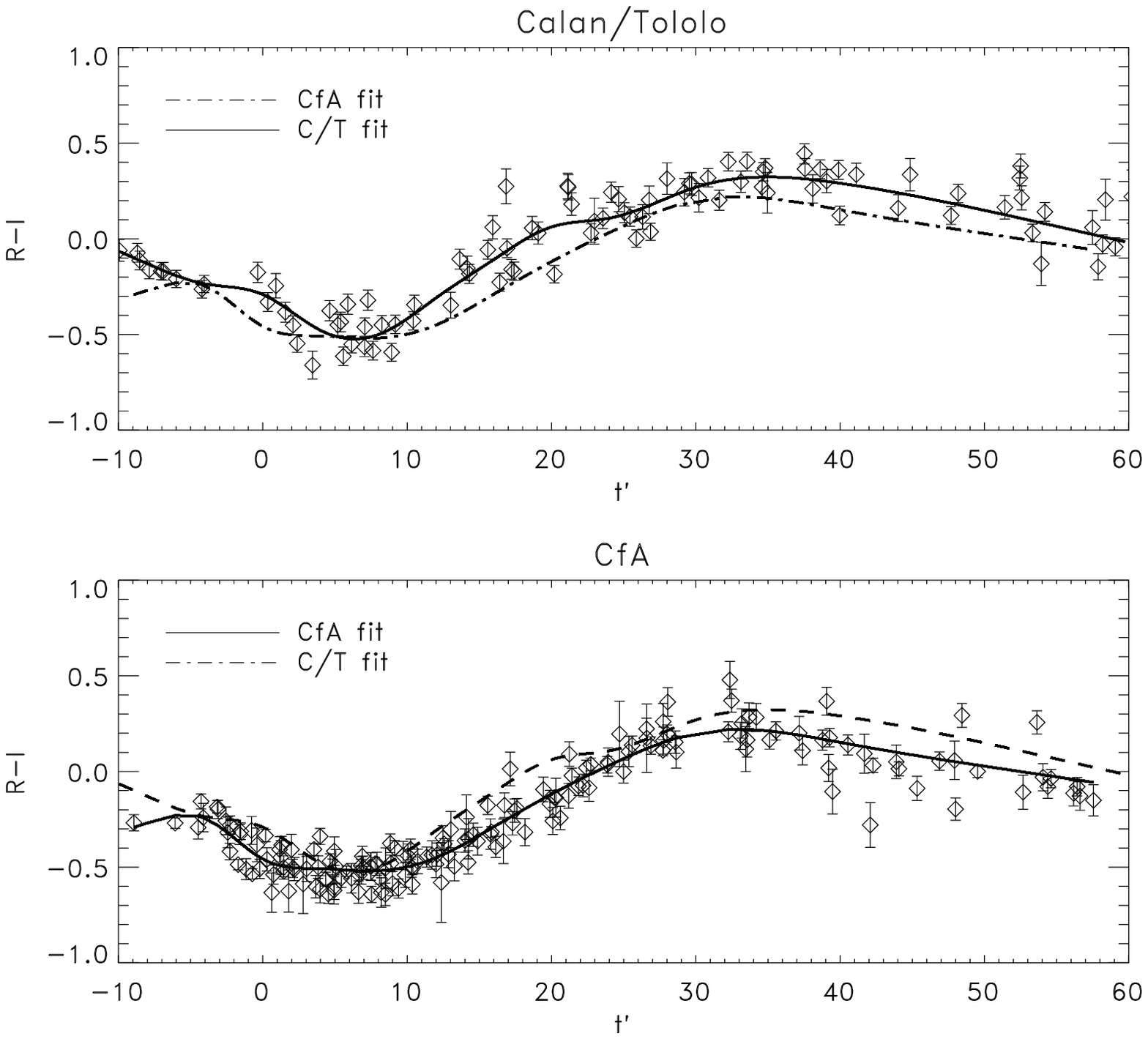}
\caption{$R-I$ for the two sets of data. The solid line represent
alway the curve found for the current set, while the dashed line is
the curve found for the same colour of the other set.}
\label{coloursRI}
\end{figure}

\begin{figure}[htb]
\includegraphics[width=8cm]{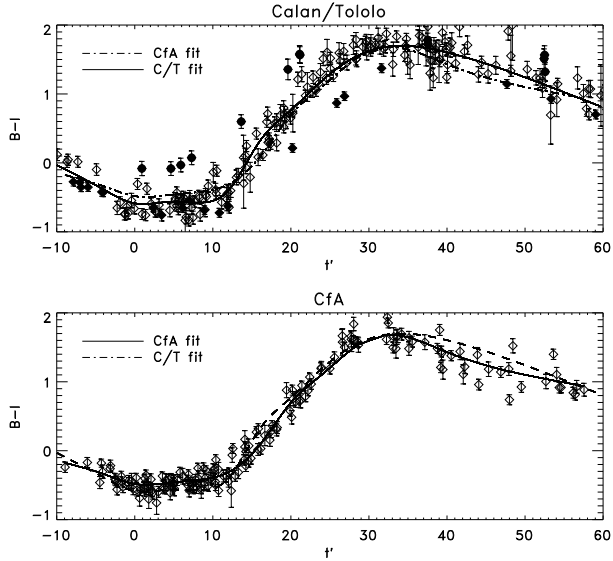}
\caption{$B-I$ for the two sets of data. The solid line represent alway
the curve found for the current set, while the dashed line is the
curve found for the same colour of the other set. The filled circles
represent SN~1993H, while the filled diamonds are data of SN~1992bc.
Refer to Sect.~\ref{sec:discussion} for a discussion about these 2
SNe.}
\label{coloursBI}
\end{figure}

\begin{figure}[htb]
\includegraphics[width=8cm]{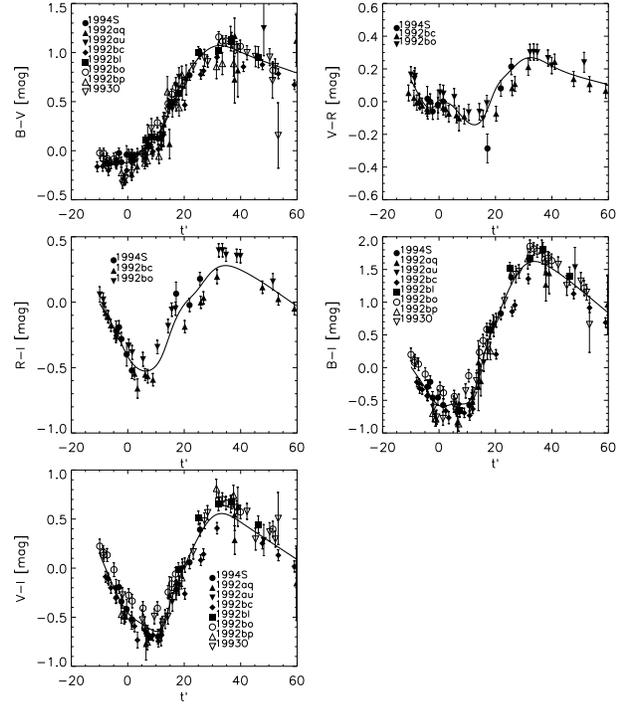}
\caption{Comparison of the models with data of unreddened SNe only.}
\label{unestincted}
\end{figure}

\begin{figure}[htb]
\includegraphics[width=8cm]{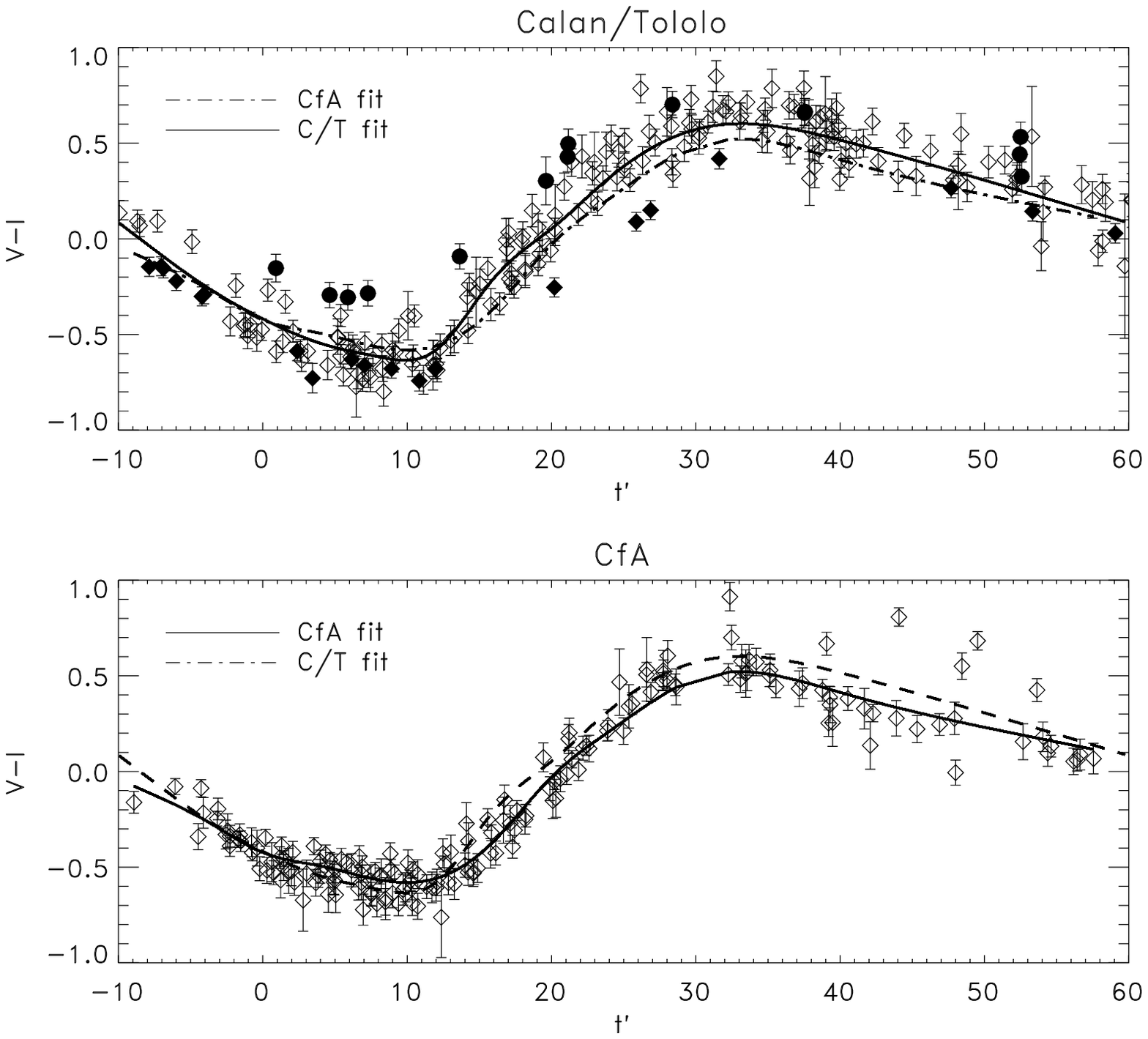}
\caption{V-I for the two sets of data. The solid line represent alway
the curve found for the current set, while the dashed line is the
curve found for the same colour of the other set. The filled circles
represent SN~1993H, while the filled diamonds are data of SN~1992bc. Refer
to Sect.~\ref{sec:discussion} for a discussion about these 2 SNe.}
\label{coloursVI}
\end{figure}

Figs.~\ref{coloursBV}-~\ref{coloursVI} show the time evolution of the
extinction and $K$-corrected $B-V$, $V-R$, $R-I$, $B-I$ and $V-I$,
colours. The plotted errors include the uncertainty on the host galaxy
exinction correction. The time axis ($t'$) has been corrected by the
SN redshift and the ${\mathit B}$-band stretch factor $s$, to account for the
dependence of colours on the stretch. This rescaling of the time axis
was found very effective in reducing the measured intrinsic dispersion
(Sect.~\ref{sec:intrinsic})
\begin{equation}
{t'}={t-t_{B_{\rm max}} \over s \cdot (1+z)}
 \label{tprime}
\end{equation}

\begin{figure}[htb]
\includegraphics[width=8.5cm]{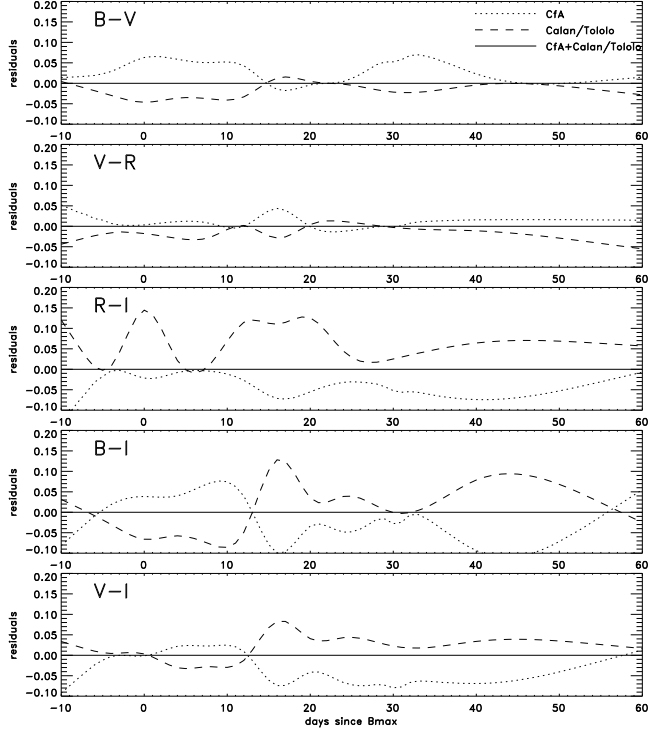}
\caption{Residuals of each of the models from the one built for all
the data set.}
\label{compare_curve}
\end{figure}

For each colour lightcurve we have spline-interpolated the weighted
mean values computed in four days wide, non-overlapping bins. The aim
of this procedure was to find a curve that describes the {\it average}
time evolution of the colours. This parameterization will be referred
as a ``{\it model}'' in the following discussion. Other methods were
used to fit the data, as for example a least squares cubic spline fit.
These result in curves that differ from the {\it model} typically by
0.01 magnitudes. The {\it models} show some systematic discrepancies
between the two data sets, especially in $R-I$ for which the C-T {\it
model} is always redder than the CfA {\it model}. In Fig.
~\ref{compare_curve} we investigate the differences of each of the {\it
models} from the one built on both sets together. The differences are
usually of the order of a few hundreds of a magnitude. The largest
deviation was found for $R-I$ in the Calan/Tololo set, resulting in a
difference of about 0.2 mag at maximum with respect to the {\it model}
built up using data from both sets. Note, however, that the statistics
in the Calan/Tololo set for $R-I$ colour, all along the evolution and
in particular around the time of  $B_{\rm max}$, is extremely poor, as
shown in Fig.~\ref{coloursRI}. Due to the smaller quoted observational
error bars, the CfA set dominates the weighted average used to build
the {\it model} out of both data sets.

A  source of uncertainty in this analysis is the ability of the
observers to convert the instrumental magnitudes from the used filter+CCD system
transmission into the standard ${\mathit BVRI}$ system. The apparent systematic
effects for $R-I$ in Fig.~\ref{compare_curve} may be indicative of this.

In order to assess systematic effects in builing the
colour {\it models}, we checked whether
the host galaxy extinction was over-corrected. Thus, we compared the {\it
models} with the data of those SNe that, according to
\citet{Phillips99}, suffered no extinction from the host galaxy. The
comparison, shown in Fig.~\ref{unestincted}, exhibits
no obvious deviations and we may conclude that the extinction corrected
colours are consistent with the uncorrected sub-sample. 

A possible remaining dependence of colours on the stretch factor $s$
was investigated. For data points in a 5 days broad bin around time of
B-maximum and around day 15, a linear weighted fit was done on the
residuals from the models not showing any further significant
dependence on $\Delta m_{\rm 15}$ or equivalently on the stretch parameter
$s$. Note that this is not in contradiction with Fig.~\ref{BmaxVmax},
as we are considering $(B-V)_{t_{B_{\rm max}}}$ instead of $B_{\rm max}-V_{\rm max}$.
Moreover, our analysis includes data points in a broad bin around day
zero, and not a fitted lightcurve maximum as in Fig.~\ref{BmaxVmax}.

\section{Intrinsic colour dispersion}
\label{sec:intrinsic}

The observed colour dispersion around the {\it models} is overall
inconsistent with a statistical random distribution of the data with
the reported measurement errors. The $\chi^2$ value per degree of
freedom computed for each model is typically between 3 and 6, giving indication
of a very poor fit of the data. Thus, we conclude that the analysed
data supports the existence of an {\it intrinsic colour dispersion} of
Type Ia supernovae \footnote{This has been further investigated as described at
the end of this section.}.

The intrinsic colour dispersion was computed on the residuals of each
of the data sets from the corresponding {\it model}, built as
explained above. Labeling $XY_i$ any of the measured colours,
$B-V$, $V-R$, $R-I$, $B-I$ and $V-I$, the residual with respect to the model
expectation is referred as $R_{{XY}_i}$, i.e.,
\begin{equation}
R_{{XY}_i}={XY}_i-XY^{\rm model}_i
\label{residuals}
\end{equation}
Next, the time axis was binned into 5 days wide bins to
compensate for the modest statistics.  The bins are centered at days
$d = 5 \cdot n$, with $n$ ranging from 0 to 8.  For each bin $k$, we
compute the weighted average for the data ${XY}$, (see Appendix~\ref{app})
\begin{equation}
XY^k = \frac{\sum_{i=1}^{N_k} w_i XY_i}{\sum_{i=1}^{N_k} w_i}
\label{weighted_average}
\end{equation}
\noindent where $N_k$ is the number of points in bin $k$ and $w_i$ is
the inverse of the uncertainty on the $i$-th measurement squared,
$1/\sigma_i^2$, where $\sigma_i$ include both measurement errors and
uncertainty on the host galaxy extinction corrections. The weighted
sample standard deviation was computed as the square root of the
weighted second moment:

\begin{equation}
s_{XY}^k=\sqrt{m_{w_2}}=\sqrt{\frac{\sum_{i=1}^N w_i R_{{XY}_i}^2}{\sum_{i=1}^N{w_i}}}
\label{disp}
\end{equation}

The uncertainty on the expression in Eq.~(\ref{disp}) is given by
the square root of the variance, $V[s_{XY}^k]$, computed as:

\begin{equation}
V[s_{XY}^k]=\frac{V[m_{w_2}]}{4m_{w_2}}
\label{edisp}
\end{equation}

\begin{table}[!htb]\caption[table data]{Results of the analysis of all SNe. First column:
central value in days for each time bin; $N_k$ is the number of points
for each bin; $XY^k$ is the weighted mean value and its 1-sigma
uncertainty; $\Delta_{XY}$ is the intrinsic dispersion computed
according to Eqs.~(\ref{disp}) and~(\ref{edisp}); $\Delta_{XY}^{\rm
corr}$ is the corrected intrinsic dispersion, computed as in
Eq.~(\ref{disp_b}), and in the last column is the estimated lower limit
at 99\% C.L. (see text); \\ $^{a}$~compatible with null intrinsic
dispersion at 99\% C.L.; an upper limit is given instead of the
corrected intrinsic dispersion. }
\begin{center}
\begin{tabular}{llccll}
\hline
\hline
day & $N_k$ & $BV^k$  &  $\Delta_{BV}$ & $\Delta_{BV}^{\rm corr}$ & L.L. \\
\hline
   0  &  57  &   -0.11 $\pm$   0.01  &  0.09 $\pm$  0.01  &  0.07  &  0.05  \\
   5  &  70  &    0.02 $\pm$   0.01  &  0.10 $\pm$  0.01  &  0.08  &  0.05  \\
  10  &  75  &    0.16 $\pm$   0.01  &  0.11 $\pm$  0.01  &  0.09  &  0.06  \\
  15  &  59  &    0.47 $\pm$   0.02  &  0.11 $\pm$  0.01  &  0.09  &  0.06  \\
  20  &  47  &    0.75 $\pm$   0.03  &  0.13 $\pm$  0.02  &  0.11  &  0.08  \\
  25  &  39  &    0.95 $\pm$   0.02  &  0.10 $\pm$  0.02  &  0.08  &  0.06  \\
  30  &  34  &    1.08 $\pm$   0.02  &  0.11 $\pm$  0.02  &  0.09  &  0.07  \\
  35$^{a}$   &  27  &    1.07 $\pm$   0.02  &  0.08 $\pm$  0.03  & $<$ 0.05 &  \\
  40  &  35  &    1.02 $\pm$   0.02  &  0.10 $\pm$  0.03  &  0.07  &  0.06  \\
\hline \hline
day & $N_k$ & $VR^k$ & $\Delta_{VR}$ & $\Delta_{VR}^{\rm corr}$ & L.L. \\ \hline
   0  &  31  &    0.04 $\pm$   0.02  &  0.07 $\pm$  0.01  &  0.06  &  0.05  \\
   5  &  36  &   -0.01 $\pm$   0.02  &  0.08 $\pm$  0.01  &  0.06  &  0.05  \\
  10  &  38  &   -0.08 $\pm$   0.01  &  0.06 $\pm$  0.01  &  0.04  &  0.04  \\
  15  &  34  &   -0.07 $\pm$   0.02  &  0.07 $\pm$  0.02  &  0.06  &  0.05  \\
  20  &  24  &    0.09 $\pm$   0.02  &  0.06 $\pm$  0.02  &  0.05  &  0.04  \\
  25$^{a}$  &  22  &    0.22 $\pm$   0.02  &  0.06 $\pm$  0.01  &  $<$ 0.04 & \\
  30$^{a}$  &  18  &    0.31 $\pm$   0.01  &  0.06 $\pm$  0.01  &  $<$ 0.04 &  \\
  35$^{a}$  &  16  &    0.30 $\pm$   0.01  &  0.04 $\pm$  0.01  &  $<$ 0.03 & \\
  40$^{a}$  &  19  &    0.25 $\pm$   0.01  &  0.05 $\pm$  0.01  &  $<$ 0.03 &  \\
\hline \hline
day & $N_k$ & $RI^k$ & $\Delta_{RI}$ & $\Delta_{RI}^{\rm corr}$ & L.L.\\ \hline
   0  &  30  &   -0.44 $\pm$   0.02  &  0.10 $\pm$  0.01  &  0.09  &  0.06  \\
   5  &  36  &   -0.50 $\pm$   0.02  &  0.09 $\pm$  0.01  &  0.07  &  0.05  \\
  10  &  32  &   -0.50 $\pm$   0.02  &  0.07 $\pm$  0.01  &  0.05  &  0.05  \\
  15  &  30  &   -0.25 $\pm$   0.04  &  0.12 $\pm$  0.03  &  0.11  &  0.08  \\
  20  &  20  &   -0.06 $\pm$   0.04  &  0.13 $\pm$  0.02  &  0.12  &  0.09  \\
  25  &  22  &    0.09 $\pm$   0.02  &  0.08 $\pm$  0.01  &  0.05  &  0.05  \\
  30  &  18  &    0.23 $\pm$   0.03  &  0.08 $\pm$  0.01  &  0.06  &  0.06  \\
  35$^{a}$  &  16  &    0.27 $\pm$   0.03  &  0.09 $\pm$  0.01  &  $<$ 0.06 &  \\
  40  &  17  &    0.22 $\pm$   0.04  &  0.14 $\pm$  0.03  &  0.13  &  0.10  \\
\hline \hline
day & $N_k$ & $BI^k$ & $\Delta_{BI}$ & $\Delta_{BI}^{\rm corr}$&
L.L.\\ \hline
   0  &  39  &   -0.54 $\pm$   0.02  &  0.13 $\pm$  0.01  &  0.10  &  0.08  \\
   5  &  48  &   -0.52 $\pm$   0.02  &  0.15 $\pm$  0.02  &  0.13  &  0.09  \\
  10  &  48  &   -0.44 $\pm$   0.03  &  0.17 $\pm$  0.01  &  0.15  &  0.10  \\
  15  &  35  &    0.10 $\pm$   0.05  &  0.18 $\pm$  0.02  &  0.16  &  0.11  \\
  20  &  34  &    0.74 $\pm$   0.06  &  0.24 $\pm$  0.05  &  0.23  &  0.15  \\
  25  &  25  &    1.25 $\pm$   0.05  &  0.22 $\pm$  0.04  &  0.21  &  0.15  \\
  30  &  27  &    1.66 $\pm$   0.03  &  0.16 $\pm$  0.02  &  0.13  &  0.10  \\
  35$^{a}$  &  23  &    1.65 $\pm$   0.03  &  0.12 $\pm$  0.02  &  $<$ 0.08 &  \\
  40  &  33  &    1.51 $\pm$   0.04  &  0.20 $\pm$  0.02  &  0.17  &  0.13  \\
\hline \hline
day & $N_k$ & $VI^k$ &$\Delta_{VI}$ & $\Delta_{VI}^{\rm corr}$& L.L.\\
\hline
   0  &  37  &   -0.43 $\pm$   0.02  &  0.08 $\pm$  0.01  &  0.05  &  0.05  \\
   5  &  51  &   -0.54 $\pm$   0.02  &  0.11 $\pm$  0.01  &  0.09  &  0.06  \\
  10  &  48  &   -0.60 $\pm$   0.01  &  0.09 $\pm$  0.01  &  0.06  &  0.05  \\
  15  &  38  &   -0.34 $\pm$   0.03  &  0.13 $\pm$  0.02  &  0.11  &  0.08  \\
  20  &  36  &    0.01 $\pm$   0.03  &  0.14 $\pm$  0.02  &  0.13  &  0.09  \\
  25  &  28  &    0.33 $\pm$   0.03  &  0.15 $\pm$  0.02  &  0.14  &  0.10  \\
  30  &  28  &    0.57 $\pm$   0.03  &  0.12 $\pm$  0.02  &  0.10  &  0.08  \\
  35  &  25  &    0.58 $\pm$   0.02  &  0.10 $\pm$  0.01  &  0.08  &  0.07  \\
  40  &  33  &    0.48 $\pm$   0.03  &  0.13 $\pm$  0.01  &  0.11  &  0.08  \\
\hline 
\end{tabular}
\label{histotable_all}
\end{center}
\end{table}

\noindent where $V[m_{w_2}]$ is the variance of the weighted second
moment. We can consider the result of Eq.~(\ref{disp}) as an estimate
of the intrinsic dispersion in each bin, for each $XY$ colour:
$\Delta_{XY}=s_{XY}^k$. This has been applied at each one of the sets
separately and at the whole set of data. The results of the method for
the whole set of data are given in Table~\ref{histotable_all}.

\begin{figure}[htb]
\includegraphics[width=8.5cm]{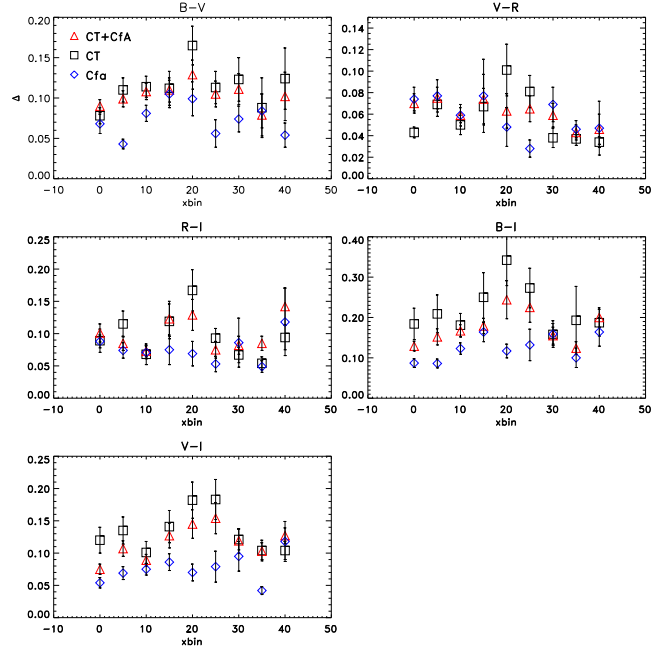}
\caption{Comparison of the results for the intrinsic dispersion in
each colour given in Table~\ref{histotable_all} for both data sets (triangles)
and the results obtained keeping the 2 sets of data separated (squares
and diamonds for the C-T and CfA respectively)}
\label{comp_intr}
\end{figure}

Fig.~\ref{comp_intr} shows a comparison of the intrinsic dispersion
computed for each of the sets separately and their combination. In
most of the cases the differences exceed the statistical
uncertainties, seemingly arising from systematic differences between
the two data sets. The computed intrinsic dispersion for the CfA
sample was found to be smaller than for the CT data set and the whole
(combined) sample in most cases, again pointing at systematic 
differences in the reported magnitudes.

Note that the adopted method leads to an overestimation of the
intrinsic colour dispersion due to the contribution from the
measurement errors. However, the weighting procedure ensures that the
most accurate measurements dominate the result. To asses the impact of
the measurement accuracy, we run a Monte-Carlo simulation to generate
a synthetic colour data set with a dispersion given by the measurement
uncertainties alone, i.e. no {\em intrinsic} dispersion. Three hundred
data sets, with the same distributions in epochs and formal error bars
as the CfA and Calan/Tololo were simulated and the weighted standard
deviation (and its error) were computed, according to
Eqs.~(\ref{disp}) and~(\ref{edisp}). Note that the simulation, for
simplicity, generates gaussian distributed and completely uncorrelated
data. The averages, $\delta$, were used to disentangle the
contribution of the intrinsic dispersion from the measurement errors.
First, an hypothesis test was run to verify whether the simulated data
and the measured data had the same dispersion; e.g. implying null
intrinsic dispersion:

\begin{tabular}{l}
$H_0: \Delta=\delta$\\
$H_1: \Delta \neq \delta$\\
\end{tabular}

A level of significance $\alpha=0.01$ was set for rejecting the null
hypothesis, \citep{Cowan1998}. Only 10 cases were not rejected,
indicated by a $^{(a)}$ in Table~\ref{histotable_all}. For all the
other cases, for which the $H_0$ hypothesis was rejected, the
intrinsic dispersion was computed as:
\begin{equation}
\Delta^{\rm corr}=\sqrt{\Delta^2-\delta^2}\\
\label{disp_b}
\end{equation}

\noindent and a lower limit on its value was set at a 99\% confidence
limit. The cases for which the null hypothesis was not rejected, were
considered as compatible with {\em no intrisic dispersion}, and an
upper limit on its value was set at a 99\% confidence level. The
corrected intrinsic dispersions are listed in the 5:th column of Table
~\ref{histotable_all}, together with upper and lower limits. We notice
that the narrowest colour dispersion happens for $V-R$, especially at
late times. At 25 days after $B_{\rm max}$ and later, this colour is
compatible with no intrinsic spread at all. Further, it should be
noted that at day 35, all the colours but $V-I$ are consistent with
vanishing intrinsic dispersion.

\section{Correlation}
\label{sec:correl}

The correlation between optical colours at different epochs was also
estimated. The property that was tested is whether a supernova that is
blue at a certain epoch for example, say at maximum, stays blue at all
epochs. In \citet{riess99q}, the authors argue that data measurements
more than 3 days apart may be considered as uncorrelated estimators of
colour. Our analysis does not support that assumption.\footnote{For
high-z objects this is even more questionable when some of the main
sources of uncertainty is the subtraction of a common image of the
host galaxy and the $K$-corrections.} We find significant correlations
for data points up to a month apart, as shown below. 
The method followed is essentially the one used to compute the
intrinsic dispersion. One can summarize the following steps:

\begin{itemize}
 \item Bin the data in time 
 \item Select only the SNe present in all the time bins
 \item For each bin compute the weighted average of the
 measurements belonging to the same SN.
 \item Compute the linear correlation coefficient between bins as in
 equation~\ref{corr}.
 \item Test the correlation coefficient significance (Appendix~\ref{appB})
\end{itemize}

\begin{figure*}[htb]
\includegraphics[width=17cm]{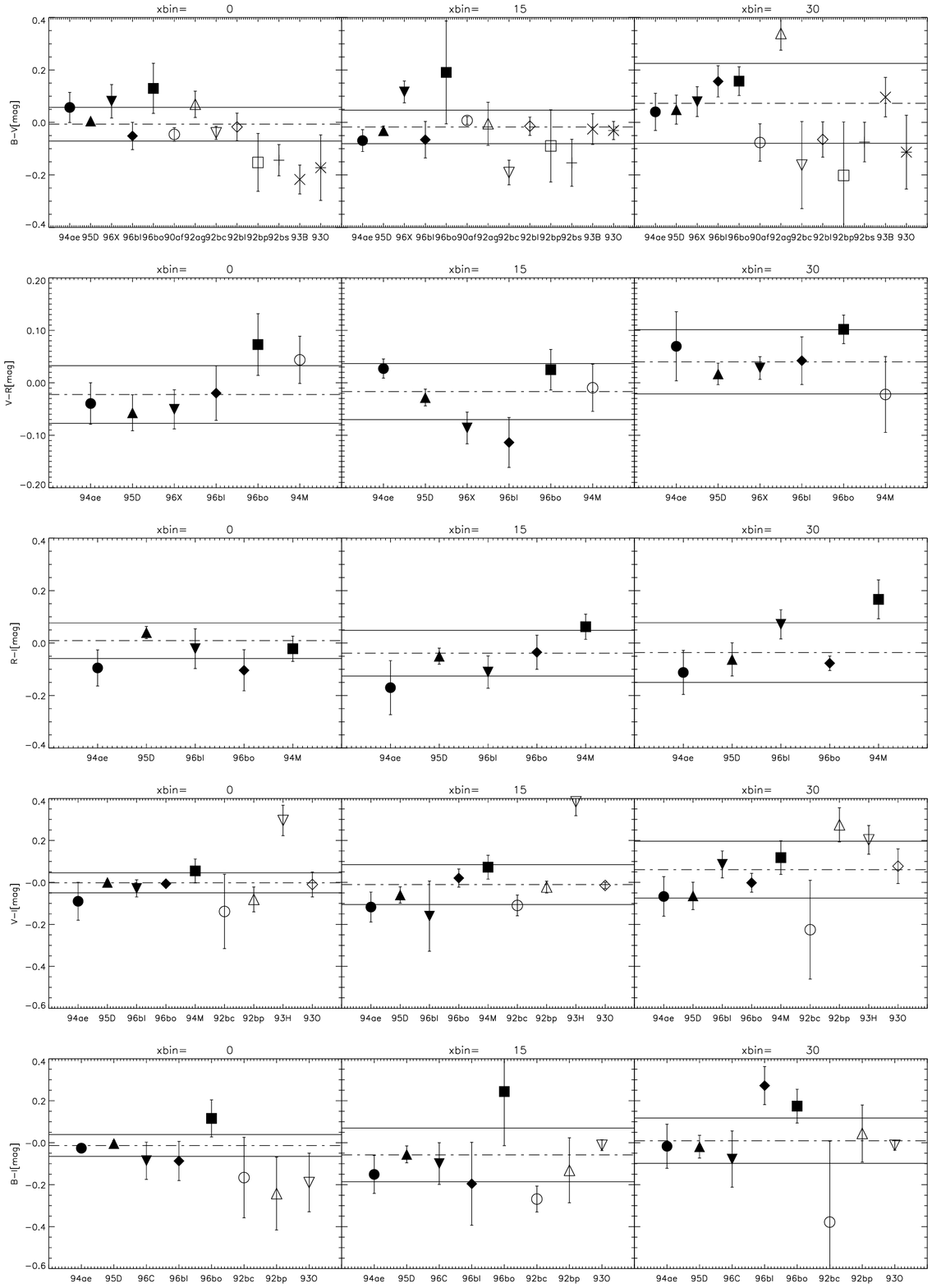}
\caption{From top to bottom and left to right. Residuals from the
models for the SNe selected for each colour, in bin centered at day
0,15 and 30. The band defined by the solid lines correspond to the
intrinsic dispersion found in at the same epoch.}
\label{corrfigure}
\end{figure*}

We refer to appendix~\ref{appB} for what follows.  The correlation
coefficient between different epochs {\it h} and {\it k} is:

\begin{equation}
r_{hk}=\frac{\sum_{i=1}^n(R_{XY_i}^h-\overline{R}_{XY}^h)(R_{XY_i}^k-\overline{R}_{XY}^k)}{\sqrt{\sum_{i=1}^n(R_{XY_i}^h-\overline{R}_{XY}^h)^2
\sum_{j=1}^n(R_{XY_j}^k-\overline{R}_{XY}^k)^2}}
\label{corr0}
\end{equation}

\noindent where the summation is on the {\it i-th} SN, which by
construction is present in all the bins.
The uncertainty on $r_{hk}$ has been computed converting it into the
normally distributed variable {\it z}, as described in (Appendix
~\ref{appB}). The results, given in Table~\ref{correl_all}, show that
the correlation is important and non-zero all along the time
evolution. Fig.~\ref{corrfigure} shows the weighted mean colours for
the selected SNe in the bins centered at day 0, 15 and 30.
Note that the supernovae selected are different in different colours,
but, by construction, are the same in all the bins for each of the
colour. It appears that supernovae that deviate from the average
colour at a certain epoch are likely to keep their colour excess all
along the 30 days evolution considered here.

\begin{table*}[htb]\caption{Correlation coefficients between the different bins. The
indicated {\it xbin} is the central value of each bin. The bin size is
7 days for all the bins. The errors indicate the 1 $\sigma$ confidence level
for the computed coefficients. See text for details.}
\begin{center}
\begin{tabular}{cccccc}
\hline\hline
{\bf $B-V$}/xbin    &      0        &    7       &      15     &   22     &     30     \\
\hline
0 & 1.00 & $  0.80_{-0.14}^{+0.09}$ & $  0.61_{-0.24}^{+0.16}$ & $
0.47_{-0.28}^{+0.21}$ & $  0.45_{-0.28}^{+0.21}$\\
7 &  & $  1.00$ & $  0.75_{-0.17}^{+0.11}$ & $  0.69_{-0.20}^{+0.13}$
& $  0.63_{-0.23}^{+0.15}$ \\
15 & & & $1.00$ & $  0.59_{-0.24}^{+0.17}$ & $  0.45_{-0.28}^{+0.22}$\\
22 & & & & $1.00$ & $  0.56_{-0.25}^{+0.18}$\\
30 & & & & & $1.00$\\
\hline
\hline
{\bf $V-R$}/xbin    &      0        &    7       &      15     &   22     &     30     \\
\hline
0 & $  1.00$ & $  0.81_{-0.31}^{+0.13}$ & $  0.46_{-0.54}^{+0.33}$ & $
0.29_{-0.56}^{+0.41}$ & $  0.03_{-0.53}^{+0.51}$\\
7 & & $  1.00$ & $  0.24_{-0.56}^{+0.44}$ & $  0.48_{-0.53}^{+0.32}$ &
$  0.24_{-0.56}^{+0.44}$\\
15 & & & $  1.00$ & $  0.48_{-0.53}^{+0.32}$ & $ -0.15_{-0.47}^{+0.55}$\\
22 & & & & $  1.00$ & $ -0.44_{-0.34}^{+0.55}$ \\
30 & & & & & $  1.00$\\
\hline
\hline
{\bf $R-I$}/xbin    &      0        &    7       &      15     &   22     &     30     \\
\hline
0 & $  1.00$ & $  0.73_{-0.51}^{+0.20}$ & $  0.57_{-0.63}^{+0.31}$ & $
0.56_{-0.63}^{+0.31}$ & $  0.31_{-0.68}^{+0.46}$\\
7 & & $  1.00$ & $  0.66_{-0.57}^{+0.25}$ & $  0.78_{-0.45}^{+0.16}$ &
$  0.58_{-0.63}^{+0.30}$\\
15 & & & $  1.00$ & $0.89_{-0.28}^{+0.08}$ & $  0.53_{-0.65}^{+0.33}$\\
22 & & & & $  1.00$ & $  0.83_{-0.38}^{+0.13}$\\
30 & & & & & $  1.00$\\
\hline
\hline
{\bf $V-I$}/xbin    &      0        &    7       &      15     &   22     &     30     \\
\hline
0 & $  1.00$ & $  0.84_{-0.17}^{+0.08}$ & $  0.92_{-0.09}^{+0.04}$ & $
0.93_{-0.08}^{+0.04}$ & $  0.49_{-0.36}^{+0.25}$\\
7 & & $  1.00$ & $  0.72_{-0.26}^{+0.15}$ & $  0.74_{-0.25}^{+0.14}$ & $  0.50_{-0.36}^{+0.24}$\\
15 & & & $  1.00$ & $  0.90_{-0.12}^{+0.06}$ & $  0.49_{-0.36}^{+0.25}$\\
22 & & & & $  1.00$ & $  0.65_{-0.30}^{+0.18}$ \\
30 & & & & & $  1.00$ \\
\hline
\hline
{\bf $B-I$}/xbin    &      0        &    7       &      15     &   22     &     30     \\
\hline
0 & $  1.00$ & $  0.62_{-0.35}^{+0.20}$ & $  0.74_{-0.27}^{+0.14}$ & $
0.44_{-0.42}^{+0.29}$ & $  0.30_{-0.44}^{+0.34}$\\
7 & & $  1.00$ & $  0.76_{-0.26}^{+0.13}$ & $  0.47_{-0.41}^{+0.27}$ & $  0.51_{-0.39}^{+0.25}$\\
15 & & & $  1.00$ & $  0.76_{-0.26}^{+0.14}$ & $  0.40_{-0.42}^{+0.30}$\\
22 & & & & $  1.00$ & $  0.73_{-0.28}^{+0.15}$\\
30 & & & & & $  1.00$\\
\hline
\end{tabular}
\label{correl_all}
\end{center}
\end{table*}

\section{Supernovae template}
\label{sec:template}
\citet{Nugent-kcorr} analysed the relation between colours of SNe~Ia
and $K$-corrections. In particular they showed how $K$-corrections are
mainly driven by the overall colour of the SN rather than by
peculiarities of single features. Using spectra and colour light
curves they give a recipe to build a {\it SN~Ia template}, to be used
for computing $K$-corrections of $s \sim 1.0$ SNe.  As the sample
used in our work is larger than the one considered in their paper, we
have used our results to improve the spectral template. Note that this
affects only the ${\mathit BVRI}$ magnitudes, which are the only bands treated
here.  Referring to \citet{Nugent-kcorr}, we proceed as follows:

\begin{itemize}

 \item correct the ${\mathit BVRI}$ of their list of ${\mathit
 UBVRIJHK}$ magnitudes using our $B-V$, $V-R$ and $R-I$ {\it models}

 \item correct the corresponding spectral templates with this
 colours\footnote{The same code used by \citet{Nugent-kcorr} was used
 for this.}

 \item iterate the whole analysis described in this paper, using the
 newly created templates to compute the $K$-corrections for both
 Calan/Tololo and CfA data set

  \item correct again the templates using the most recent version of
  the {\it models}
\end{itemize}

Even though the template itself is modified considerably from the
original one, iterating once results in small corrections ($\sim$
0.001-0.005 mag) in the intrinsic dispersion in all colours at any
epoch, and a few percent in the colour {\it models}. As the correction
obtained are small and well within the given uncertainties, we do not
iterate further. The results of this analysis should {\it not} be
considered definitive and may change as SNe~Ia and their colour
evolutions are studied in even more detail. However, it can be
considered a good estimate of the average SN~Ia template (in ${\mathit BVRI}$), as
it is constructed from a quite broad sample of SNe. The final
corrected template is available upon request.

\section{Discussion}
\label{sec:discussion}
Throughout our analysis all data points were treated as independent
measurements. This is particularly important for the cases with
significant host galaxy light underneath the supernova where a single
reference image was used, introducing a correlation between the data
points not considered here.

A first estimate of the intrinsic dispersion was calculated for each
data set separately and the two data sets together. A comparison of
the results shows that the colours extracted from the CfA sample have
smaller scatter indicating that the contribution from measurement
errors is not negligible. Further, in the analysis the measurements
are assumed to be Gaussian distributed and the weighted standard
deviation has been taken as an estimate of the intrinsic dispersion.
However, we noticed that 2 SNe, SN~1993H and SN~1992bc, seem to be
rather deviant in $B-I$ and $V-I$ for the C-T set, as shown in Figs.
~\ref{coloursBI} and ~\ref{coloursVI}. The effect of the ``outliers''
is particularly important aroud 20 days after B-maximum. For
comparison, we recalculated the intrinsic spread in $B-I$ and $V-I$
excluding these 2 SNe. The results are shown on
Table~\ref{newhistoBI_VI}. The agreement between the intrinsic
dispersion between the data sets improves when these 2 SNe are
excluded. We emphasize that there are systematic differences between
the Calan/Tololo and CfA data sets, and they might have been
introduced while converting from the instrumental system used to the
standard ${\mathit BVRI}$ system. 

An attempt of disentangling the intrinsic dispersion from the
contribution of the uncertainties has been done resulting in upper
limit values for the intrinsic dispersion in some cases. The most
relevant is the case of $V-R$, that seems compatible with null
intrinsic dispersion for most of the epochs at 99\% C.L. Moreover this
analysis brings out an important feature at day 35, when all the
colours but $V-I$ are consistent with zero intrinsic dispersion. This
indicates that further studies will be needed to investigate this
intriguing finding.
\begin{table*}[htb]
\caption[table data]{Results of the analysis done excluding SN~1992bc
  and SN~1993H (**) for $B-I$ and $V-I$. {\it xbin} is the central
  value in days for each bin; $XY^k$** is the weighted mean value and
  its uncertainty; $\Delta_{BI}$** is computed according to
  Eqs.~(\ref{disp}) and~(\ref{edisp}); $\Delta_{XY}^{\rm corr}$** is
  the corrected intrinsic disperion and L.L is the lower limit at 99\%
  C.L.; In the last column is the difference between the results given
  in Table ~\ref{histotable_all} and this analysis; the errors are
  computed as sum in quadrature. }
\begin{center}
\begin{tabular}{cccllr}
\hline\hline
day & $BI^k$**  & $\Delta_{BI}$** & $\Delta_{BI}^{\rm corr}$** & L.L.&
$\Delta_{BI}-\Delta_{BI}$**\\ 
\hline
  0  & -0.53 $\pm$  0.02  &  0.13 $\pm$  0.01  &  0.10  &  0.08  & 0.000 $\pm$ 0.017 \\
  5  & -0.53 $\pm$  0.02  &  0.13 $\pm$  0.01  &  0.10  &  0.08  & 0.024 $\pm$ 0.025 \\
 10  & -0.41 $\pm$  0.02  &  0.15 $\pm$  0.02  &  0.13  &  0.09  & 0.018 $\pm$ 0.022 \\
 15  &  0.10 $\pm$  0.05  &  0.18 $\pm$  0.02  &  0.16  &  0.11  & 0.003 $\pm$ 0.026 \\
 20  &  0.74 $\pm$  0.05  &  0.15 $\pm$  0.02  &  0.12  &  0.09  & 0.098 $\pm$ 0.052 \\
 25  &  1.31 $\pm$  0.05  &  0.12 $\pm$  0.02  &  0.09  &  0.08  & 0.101 $\pm$ 0.044 \\
 30  &  1.68 $\pm$  0.03  &  0.12 $\pm$  0.02  &  0.08  &  0.08  & 0.038 $\pm$ 0.028 \\
 35  &  1.65 $\pm$  0.03  &  0.13 $\pm$  0.02  & $<$ 0.09 &  &-0.006 $\pm$ 0.024 \\
 40  &  1.49 $\pm$  0.04  &  0.20 $\pm$  0.03  &  0.18  &  0.13  &-0.004 $\pm$ 0.035 \\
\hline
day & $VI^k$** & $\Delta_{VI}$** & $\Delta_{VI}^{\rm corr}$** & L.L.  & $\Delta_{VI}-\Delta_{VI}$**\\ 
\hline
  0  & -0.43 $\pm$  0.02  &  0.07 $\pm$  0.01  &  0.05  &  0.05  & 0.002 $\pm$ 0.011 \\
  5  & -0.55 $\pm$  0.01  &  0.09 $\pm$  0.01  &  0.06  &  0.05  & 0.019 $\pm$ 0.014 \\
 10  & -0.58 $\pm$  0.01  &  0.09 $\pm$  0.01  &  0.06  &  0.05  & 0.002 $\pm$ 0.013 \\
 15  & -0.34 $\pm$  0.03  &  0.11 $\pm$  0.01  &  0.09  &  0.07  & 0.016 $\pm$ 0.022 \\
 20  &  0.01 $\pm$  0.03  &  0.11 $\pm$  0.02  &  0.09  &  0.07  & 0.036 $\pm$ 0.027 \\
 25  &  0.35 $\pm$  0.03  &  0.09 $\pm$  0.01  &  0.07  &  0.06  & 0.060 $\pm$ 0.028 \\
 30  &  0.57 $\pm$  0.03  &  0.11 $\pm$  0.02  &  0.09  &  0.07  & 0.007 $\pm$ 0.029 \\
 35  &  0.58 $\pm$  0.02  &  0.10 $\pm$  0.01  &  0.08  &  0.07  & 0.000 $\pm$ 0.019 \\
 40  &  0.47 $\pm$  0.03  &  0.13 $\pm$  0.01  &  0.11  &  0.08  & 0.000 $\pm$ 0.018 \\
\hline
\end{tabular}
\label{newhistoBI_VI}
\end{center}
\end{table*}
When computing the correlation coefficients we considered data up to
30 days after ${\mathit B}$-band maximum, even though later epoch data are
available for several supernovae. However, the necessary condition of
each SN being observed in all the time bins reduces the statistics if
later epochs are introduced. This limitation is specific to the sample
used and can be overcome with the use of a more extensively observed
sample, such as what will be provide by SNfactory \citep{SNfactory}.

The intrinsic dispersion sets constrains on the ability to determine
the host galaxy extinction,$A_V$. This will depend on the colour, as
the intrinsic dispersion is different for different colours. As
an example we used the \citet{Cardelli89} relation at the effective
wavelength for each bandpass to compute the expected uncertainty in
the extinction, neglecting any dependence on the supernova phase.

\begin{equation}
\begin{array}{lll}
\sigma_{A_V}^{\rm B-V}& = & 3.1\cdot \Delta_{\rm BV}^{\rm corr}\\
\sigma_{A_V}^{\rm V-R}& = & 6.2\cdot \Delta_{\rm VR}^{\rm corr}\\
\sigma_{A_V}^{\rm R-I}& = & 4.1\cdot \Delta_{\rm RI}^{\rm corr}\\
\sigma_{A_V}^{\rm B-I}& = & 1.4\cdot \Delta_{\rm BI}^{\rm corr}\\
\sigma_{A_V}^{\rm V-I}& = & 2.5\cdot \Delta_{\rm VI}^{\rm corr}\\
\end{array}
\label{eq:sigma_ext}
\end{equation}
where $R_V$ was assumed equal to 3.1. Table~\ref{tab:sigma_ext} shows
the results of the equation~\ref{eq:sigma_ext} for the epochs for
which the intrinsic dispersion was calculated. The $\Delta_{XY}^{\rm
corr}$ was used for this. The results indicate that, with the present
knowledge, extinction by dust with $R_V=3.1$ may only be determined to
$\sigma_{A_V} \gsim 0.10$ with Type Ia restframe optical data within
the first 40 days after ${\mathit B}$-band lightcurve maximum, for the
colours and epochs with non-zero intrinsic dispersion. However
observations in $V-R$ are preferable to other colours to set limit on
the extinction of an observed supernova. To account for 
possible different extinction parameters $R_V$ in the 
different supernova host galaxies (as noticed in e.g.
\citep{RiessPressKirshner96} and \citep{Krisciunas2000}), we repeated
the analysis considering a Gaussian uncertainty $\sigma_{R_V}=1$ on $R_V$, 
propagating this scatter on the host galaxy extinction
corrections. This changes the values of the intrinsic 
dispersion, but typically within the quoted errors in
Table~\ref{histotable_all}.

\begin{table}[htb]
\caption{$\sigma_{A_V}$ represent the constrain on the extinction
$A_V$ that it is possible to compute for a given intrinsic dispersion.
The $\Delta_{XY}^{\rm corr}$ and it L.L given in
Table~\ref{histotable_all} has been used for these estimates.}
\begin{center}
\begin{tabular}{cccc}
\hline\hline
 & xbin & $\sigma_{A_V}$ & L.L.\\
\hline
$B-V$ &    0 &   0.23  &  0.16 \\
$B-V$ &    5 &   0.25  &  0.17 \\
$B-V$ &   10 &   0.28  &  0.18 \\
$B-V$ &   15 &   0.28  &  0.20 \\
$B-V$ &   20 &   0.34  &  0.24 \\
$B-V$ &   25 &   0.25  &  0.20 \\
$B-V$ &   30 &   0.27  &  0.22 \\
$B-V$ &   35 &   $<$  0.16 & \\
$B-V$ &   40 &   0.21  &  0.20 \\
\hline
$V-R$ &    0 &   0.36  &  0.28 \\
$V-R$ &    5 &   0.38  &  0.29 \\
$V-R$ &   10 &   0.27  &  0.22 \\
$V-R$ &   15 &   0.36  &  0.29 \\
$V-R$ &   20 &   0.29  &  0.26 \\
$V-R$ &   25 &   $<$  0.27 & \\
$V-R$ &   30 &   $<$  0.25 & \\
$V-R$ &   35 &   $<$  0.19 & \\
$V-R$ &   40 &   $<$  0.20 & \\
\hline
$R-I$ &    0 &   0.37  &  0.27 \\
$R-I$ &    5 &   0.28  &  0.22 \\
$R-I$ &   10 &   0.22  &  0.19 \\
$R-I$ &   15 &   0.46  &  0.32 \\
$R-I$ &   20 &   0.48  &  0.36 \\
$R-I$ &   25 &   0.22  &  0.21 \\
$R-I$ &   30 &   0.25  &  0.23 \\
$R-I$ &   35 &   $<$  0.25 & \\
$R-I$ &   40 &   0.54  &  0.41 \\
\hline
$V-I$ &    0 &   0.12  &  0.12 \\
$V-I$ &    5 &   0.22  &  0.16 \\
$V-I$ &   10 &   0.16  &  0.13 \\
$V-I$ &   15 &   0.27  &  0.20 \\
$V-I$ &   20 &   0.32  &  0.22 \\
$V-I$ &   25 &   0.35  &  0.25 \\
$V-I$ &   30 &   0.25  &  0.19 \\
$V-I$ &   35 &   0.19  &  0.17 \\
$V-I$ &   40 &   0.26  &  0.20 \\
\hline
$B-I$ &    0 &   0.14  &  0.11 \\
$B-I$ &    5 &   0.18  &  0.13 \\
$B-I$ &   10 &   0.20  &  0.14 \\
$B-I$ &   15 &   0.22  &  0.16 \\
$B-I$ &   20 &   0.32  &  0.21 \\
$B-I$ &   25 &   0.29  &  0.21 \\
$B-I$ &   30 &   0.18  &  0.14 \\
$B-I$ &   35 & $<$  0.12 & \\
$B-I$ &   40 &   0.24  &  0.18 \\
\hline
\end{tabular}
\label{tab:sigma_ext}
\end{center}
\end{table}

\section{Conclusion}
\label{sec:concl}

A statistical analysis of colours of SNe~Ia using a sample of 48
nearby SNe was performed. Of this sample 7 SNe have been excluded
based on their large extinction, or peculiar behavior of their colours
at maximum. With the present knowledge and data quality we computed
the average colour evolution for $B-V$, $V-R$, $R-I$, $B-I$ and $V-I$ with time
and the derived intrinsic scatter. We find that the correlation of 
colours during the first 30 days after restframe ${\mathit B}$-band maximum is 
not negligible, i.e. arguing against the assumptions made in the 
analysis of SN~1999Q \citep[z=0.46;][]{riess99q} where
five measurements of restframe $B-I$ along its lightcurve were treated
as {\em independent} estimates of extinction. According to our 
findings, their limit on the presence of intergalactic grey dust must
be revised. A reanalysis of the data from SN~1999Q is in preparation
\citep{beethoven}.
With the data at hand, host galaxy extinction corrections from restframe 
optical colours within the first 30 days after maximum light are 
generaly limited to $\sigma_{A_V} \gsim 0.1$ due to the intrinsic 
variation of Type Ia colours at those epochs, with the possible 
exception of extinction corrections derived from the rest-frame
$V-R$ colour. 
The results of this analysis have been used to correct spectroscopic
templates which are available upon request.

\begin{acknowledgements}
S. Nobili wishes to thank Christian Walck for his valuable help and
constructive discussions on the statistical analysis presented in this
paper and Don Groom, Chris Lidman and Vallery Stanishev for their
careful reading and useful suggestions. SN is supported by a graduate
student grant from the Swedish Research Council. AG is a Royal Swedish
Academy Research Fellow supported by a grant from the Knut and Alice
Wallenberg Foundation. PEN acknowledges support from a NASA LTSA grant
and by the Director, Office of Science under U.S. Department of Energy
Contract No. DE-AC03-76SF00098.
\end{acknowledgements}

\appendix

\renewcommand{\theequation}{A-\arabic{equation}}
\setcounter{equation}{0}  
\section{Appendix}  
\label{app}

\subsection{Algebraic moments}
\label{appA}

In what follows we refer to \citet{Cowan1998} and
\citet{Kendall58} (see Volume 1, Chapter 10),
our aim here is to give uniformity to the notations used. Given n
independent observations of a variable, $x_1, x_2, ..., x_n$, the r-th
moment or algebraic moment is given by:
\begin{equation}
m_r=\frac{1}{n}\sum_{i=1}^n x_i^r
\label{m}
\end{equation}
\noindent The expectation value, or mean value, and the variance of $m_r$ are:

\begin{equation}
\begin{tabular}{l}
$E[m_r]=\mu_r$\\
$V[m_r]=\frac{1}{n}(\mu_{2r}-\mu_r^2)$\\
\end{tabular}
\label{Em}
\end{equation}

\noindent where $\mu_{2r}$ is the $2\cdot r$-th moment. The ~\ref{Em} are exact
formulae as long as one knows $\mu_{2r}$ and $\mu_r$. However this is
not always the case, and one has to use their estimators $m_{2r}$ and
$m_r$ from the sample itself. The variance of the standard deviation
can be computed using standard error propagation:

\begin{equation}
V[\sigma]=V[\sqrt{m_2}]=\frac{V[m_2]}{4m_2}=\frac{m_4-m_2^2}{4nm_2}
\label{Vsigma}
\end{equation}

In experimental situation each observation $x_i$ is often
attached to a certain weight $w_i$. Supposing that the weight
themselves are known without errors, one can define the following
formula for the r-th weighted moments:

\begin{equation}
m_{w_r}=\frac{\sum_{i=1}^n w_ix_i^r}{\sum_{i=1}^n w_i}
\label{weigthedall}
\end{equation}

\noindent In order to evaluate the variance on the weighted standard deviation
we simply extended the ~\ref{Vsigma} for the case of weighted moments,
obtaining:

\begin{equation}
V[\sigma_w]=V[\sqrt{m_{w_2}}]=\frac{V[m_{w_2}]}{4m_{w_2}}=\frac{m_{w_4}-m_{w_2}^2}{4n_{eff}m_{w_2}}
\label{Vsigmaw}
\end{equation}

\noindent where the $m_{w_4}$ and $m_{w_2}$ are the 4th and 2nd weighted
moments respectively and $n_{eff}=(\sum_{i=1}^nw_i)^2/\sum_{i=1}^nw_i^2$.


\noindent A simple Monte Carlo simulation has been run to verify the accuracy of
the approximated formula~\ref{Vsigmaw}.

\subsection{Correlation coefficient}
\label{appB}

Given 2 random variables {\it x} and {\it y} the correlation coefficient is
defined as

\begin{equation}
  \rho_{xy}=\frac{V_{xy}}{\sigma_x \sigma_y}
\end{equation}

The unbiased estimator of the covariance $V_{xy}$ is:

\begin{equation}
  \widehat{V}_{xy}=\frac{1}{n-1}\sum_{i=1}^n(x_i-\overline{x})(y_i-\overline{y})
\end{equation}

\noindent so that the estimator of the correlation coefficient will be:

\begin{equation}
r_{xy}=\frac{\widehat{V}_{xy}}{s_x 
s_y}=\frac{\sum_{i=1}^n(x_i-\overline{x})(y_i-\overline{y})}{\sqrt{\sum_{i=1}^n(x_i-\overline{x})^2 \sum_{j=1}^n(y_j-\overline{y})^2}}
\label{corr}
\end{equation}

For the ~\ref{corr} to be a good estimator of the correlation
coefficient there are a few caveats to check. First condition is that
the samples are randomly defined from the population, that is to make
sure that the samples x and y are not selected in some ways that would
operate as to increase or decrease the value of r. Even though one has
random samples it is possible to compute the errors due to sampling.
Commonly this is computed as $(1-r^2)/\sqrt{n}$. Unfortunately this
is just an approximation. Moreover r's for successive samples are not
distributed normally unless n is large and the true value $\rho$ is
near zero.
This yields to a distinction:

\begin{itemize}
\item if $n > 30$\\ in order to know whether the value calculated for
r is significantly different from zero, one can compute its standard error as:

\begin{equation}
\sigma_r=1/\sqrt{n-1}
\end{equation}

If $r/\sigma$ is greater than 2.58, one can conclude that the universe
value of r is likely to be greater than zero. 

\item if $n < 30$\\
the variable:

\begin{equation}
t=r\frac{\sqrt{n-2}}{\sqrt{1-r^2}}
\end{equation}

follows the t-distribution with dof=n-2.
This can be used only for testing the hypothesis of zero correlation.

\end{itemize}

  R. A. Fisher developed a technique to overcome these
difficulties.  The variable {\it r} is transformed into another
variable that is normally distributed. This is especially useful for
high value of {\it r}, when none of the above test can be safely applied.
The transformation to the variable {\it z}:

\begin{equation}
z=\frac{1}{2}\ln(1+r)-\frac{1}{2}\ln(1-r)
\label{r2z}
\end{equation}

\noindent allows some important semplifications. The distribution of {\it z}'s
for successive samples does not depend on the universe value $\rho$
and the distribution of {\it z} for successive samples is so near
to normal that it can be treated as such without any loss of accuracy,
see for example (Volume 1, Chapter 16 and Volume
  2,Chapter 26) of \citet{Kendall58}. Moreover the standard error for
{\it z} is independent on its $\sigma$:

\begin{equation}
\sigma_z=1/\sqrt{n-3}
\label{sgmaz}
\end{equation}

\noindent The way to proceed is then very simple:

\begin{itemize}
\item Compute {\it r} according to (~\ref{corr})
\item Transform {\it r} into {\it z} according to (~\ref{r2z})
\item Compute $\sigma_z$ according to (~\ref{sgmaz})
\item In order to state a 1 sigma confidence limit for r, transform the 2 values 
$z \pm \sigma_z$ back to {\it r}, using the inverse of (~\ref{r2z})
\end{itemize}


\begin{thebibliography}{99}
\bibitem[Aguirre(1999a)]{aguirre1999a} Aguirre,~A., 1999, ApJ, 512, L19
\bibitem[Aguirre(1999b)]{aguirre1999b} Aguirre,~A., 1999, ApJ, 525, 583
\bibitem[Aguirre \& Haiman(2000)]{aguirre2000} Aguirre,~A. \&
  Haiman,~Z., 2000, ApJ, 532, 28.
\bibitem[Aldering et al.(2002)]{SNfactory}
Aldering,~G. et al., 2002, Supernova Factory Webpage, {\tt http://snfactory.lbl.gov}
\bibitem[Branch et al. (1993)]{bfn93} Branch,~D., Fisher,~A. \& Nugent,~P., 1993, AJ, 106, 2383
\bibitem[Cardelli et al. (1989)]{Cardelli89}Cardelli,~J.A., 
Clayton,~G.C. \& Mathis,~J.S., 1989, ApJ. 345,245
\bibitem[Cowan(1998)]{Cowan1998}Cowan,~G., 1998, Statistical data
analysis, Oxford University Press
\bibitem[Cramer(1957)]{Cramer}Cramer,~H., 1957, Mathematical methods of
statistics, Princeton University Press, Seventh Printing
\bibitem[Goldhaber et al.(2001)]{goldhaber} Goldhaber,~G.,
  Groom,~D.E., Kim,~A. et al., 2001, ApJ, 558, 359  
\bibitem[Goobar et al.(2002)]{snoc} Goobar,~A., M\"ortsell,~E., Amanullah,~R. et al., 2002 A\&A, 392, 757  
\bibitem[Hamuy et al.(1996)]{Hamuy}Hamuy,~M., Phillips,~M.M., Suntzeff,~N.B. et al., 1996,AJ, 112, 2408
\bibitem[Kendall \& Stuart (1958)]{Kendall58} Kendall,~M.G. \&
  Stuart,~A., 1958, The advanced theory of statistics, Charles Griffin \&
  Company Limited, London.
\bibitem[Krisciunas et al.(2000)]{Krisciunas2000}Krisciunas,~K.,
  Hastings,~N.C., Loomis,~K. et al., 2000, ApJ, 539, 658
\bibitem[Lira(1995)]{Lira}Lira,~P., 1995, Master Thesis, Univ. Chile
\bibitem[M\"ortsell et al. (2002)]{axion}
  M\"ortsell,~E.,Bergstr\"om,~L \& Goobar,~A., 2002, Phys.~Rev.~D, 66, 047702
\bibitem[Nobili et al.(2003)]{beethoven} Nobili,~S. et al., 2003, in prep.
\bibitem[Nugent et al.(2002)]{Nugent-kcorr}Nugent,~P., Kim,~A., \&
Perlmutter,~S., 2002, PASP, 114, 803
\bibitem[O'Donnell (1994)]{Odonnell}O'Donnell,~J.E., 1994, ApJ, 422, 158O  
\bibitem[Perlmutter et al.(1997)]{Perl97} Perlmutter,~S., Gabi,~S., Goldhaber,~G. et al., 1997, ApJ, 483,565P
\bibitem[Perlmutter et al.(1999)]{Perl99} Perlmutter.~S., Aldering,~G.,
 Goldhaber,~G. et al., 1999, ApJ, 517,565P
\bibitem[Phillips(1993)]{deltam15}Phillips,~M.M., 1993, ApJ,413L.105P
\bibitem[Phillips et al.(1999)]{Phillips99}Phillips,~M.M., Lira,~P.,
Suntzeff,~N.B., Schommer,~R.A., Hamuy,~M.\& Maza,~J., 1999, ApJ. 118, 1766-1776
\bibitem[Riess et al. (1996)]{RiessPressKirshner96}Riess,~A.G.,
  Press,~W.H., \& Kirshner,~R.P., 1996, ApJ, 473, 588R
\bibitem[Riess et al.(1998)]{Riess} Riess,~A.G., Filippenko,~A.V.,
  Challis,~P. et al., 1998, AJ, 116, 1009.
\bibitem[Riess et al.(1999)]{Riess22} Riess,~A.G., Kirshner,~R.P.,Schmidt,~B.P. et al., 1999, ApJ. 117,707-724
\bibitem[Riess et al.(2000)]{riess99q} Riess,~A.G.,Filippenko,~A.V.,
  Liu,~M.C.et al., 2000, ApJ, 536, 62  
\end{thebibliography}
\end{document}